\documentclass[aps,showpacs]{revtex4}
\usepackage{latexsym}
\usepackage{amsfonts}
\usepackage{amsmath}
\usepackage{amssymb}
\usepackage{revsymb}
\usepackage{bm}
\usepackage{graphicx}
\usepackage{axodraw}

\newcommand{\bbra}[1]{\bigl\langle #1 \bigr|}
\newcommand{\bket}[1]{\bigl| #1 \bigr\rangle}

\begin{document}

\begin{flushright}
VT-IPNAS-09-10
\end{flushright}

\title{Constraints on R-parity violation from recent Belle/Babar data}
\author{Yee Kao}\email{ykao@vt.edu}
\affiliation{Institute for Particle, Nuclear, and Astronomical Sciences,
Physics Department, Virginia Tech, Blacksburg, VA 24061}
\author{Tatsu Takeuchi}\email{takeuchi@vt.edu}
\affiliation{Institute for Particle, Nuclear, and Astronomical Sciences,
Physics Department, Virginia Tech, Blacksburg, VA 24061}

\begin{abstract}
We discuss possible constraints on R-parity violation from recently announced
Belle/Babar results on the $B\rightarrow \tau\nu$ branching fraction, and
the bounds on $\tau^-\rightarrow \ell^- K_S^0$ ($\ell=e$ or $\mu$) from Babar.
\end{abstract}


\maketitle

\section{Introduction}

Measurements of rare decay processes which are small or absent 
within the Standard Model (SM) provide windows to new physics.
During the past year, new measurements and bounds on the decays
$B\rightarrow\tau\nu_\tau$ \cite{Barrett:2009if,Hara:2008hp}
and $\tau\rightarrow\ell K^0_S$ ($\ell=e$ or $\mu$) \cite{Aubert:2009ys}
have been announced from Babar and Belle.
In this letter, we discuss what constraints can be placed on new
physics from these results using R-parity violating supersymmetry (SUSY)
as an example, 
partially updating the analyses of Dreiner et al. from 2002 \cite{Dreiner:2001kc} and 
2006 \cite{Dreiner:2006gu}.

\section{$B\rightarrow \tau\nu_\tau$}

\subsection{Experimental Value}

Babar recently reported their measurement of the 
$B \rightarrow \tau \nu_\tau$ branching fraction, using $383 \times 10^6$
$B\bar{B}$ pairs and two different methods to reconstruct the tagged $B$, 
as 
\begin{equation}
\begin{array}{ll}
\mathcal{B}(B \rightarrow \tau \nu_\tau)_{\mathrm{Babar-hadronic}} &=\;
\bigl(\,1.8 \;{}^{+0.9}_{-0.8}\,(\mathrm{stat.}) \pm 0.4\,(\mathrm{bkg.}) \pm 0.2\,(\mathrm{syst.})\,\bigr) \times 10^{-4}\;,\\ 
\mathcal{B}(B \rightarrow \tau \nu_\tau)_{\mathrm{Babar-semileptonic}} &=\;
\bigl(\,0.9 \pm 0.6\,(\mathrm{stat.}) \pm 0.1\,(\mathrm{syst.})\,\bigr) \times 10^{-4}\;,
\end{array}
\label{BtaunuBabar}
\end{equation}
with the combined value given by \cite{Barrett:2009if}
\begin{eqnarray}
\mathcal{B}(B \rightarrow \tau \nu_\tau)_{\mathrm{Babar}} \;=\; 
\bigl(\,1.2 \pm 0.4\,(\mathrm{stat.}) \pm 0.3\,(\mathrm{bkg.}) \pm 0.2\,(\mathrm{syst.})
\,\bigr) \times 10^{-4}\;. \label{BtaunuBabarCombined}
\end{eqnarray}
Belle also reported a new measurement of
$\mathcal{B}(B \rightarrow \tau \nu_\tau)$ using $657\times 10^6$ $B\bar{B}$ pairs 
and semileptonic tagging as \cite{Hara:2008hp}
\begin{equation}
\mathcal{B}(B \rightarrow \tau \nu_\tau)_{\mathrm{Belle-semileptonic}} \;=\;
\left(\,1.65\;{}^{+0.38}_{-0.37}\,(\mathrm{stat.})\;{}^{+0.35}_{-0.37}\,(\mathrm{syst.})
\,\right) \times 10^{-4}\;. 
\label{BtaunuBelleSemileptonic}
\end{equation}
A previous 2006 Belle result using $449\times 10^6$ $B\bar{B}$ pairs 
and hadronic tagging yielded \cite{Ikado:2006un}
\begin{equation}
\mathcal{B}(B \rightarrow \tau \nu_\tau)_{\mathrm{Belle-hadronic}} \;=\;
\left(\,1.79\;{}^{+0.56}_{-0.49}\,(\mathrm{stat.})\;{}^{+0.46}_{-0.51}\,(\mathrm{syst.})
\,\right) \times 10^{-4}\;.
\end{equation}
Incorporating all these results, the world average for the branching fraction 
as of August 2009 is \cite{UTfit}
\begin{equation}
\mathcal{B}(B \rightarrow \tau \nu_\tau)_{\mathrm{exp}} \;=\; (\,1.51 \pm 0.33\,) \times 10^{-4}\;.
\label{BtaunuExp}
\end{equation}
%

\subsection{Standard Model Value}

In the Standard Model (SM), the decay 
$B\rightarrow \tau\nu_\tau$ proceeds via $s$-channel $W$-exchange which
at low energies is described by the effective Lagrangian
\begin{equation}
\mathcal{L}_W \;=\;
-\sqrt{2}\,G_F V_{ub}
\bigl(\,\overline{u}\gamma^\mu(1-\gamma_5)b \,\bigr)
\bigl(\,\overline{\tau}_L\gamma_\mu \nu_{\tau\!L} \,\bigr)
\;+\; h.c.\;.
\end{equation}
This leads to the SM prediction
\begin{equation}
\mathcal{B}(B\rightarrow\tau\nu_\tau)
\;=\;\Bigl|\sqrt{2}\,G_F V_{ub} m_\tau \Bigr|^2\,\dfrac{m_B}{16\pi}
\left(1-\dfrac{m_\tau^2}{m_B^2}\right)^2 f_B^2 \tau_B\;,
\label{SMprediction}
\end{equation}
where the $B$-decay constant $f_B$ is normalized as
\begin{equation}
\bbra{0}\,\bar{u}(x)\gamma^\mu\gamma_5 b(x)\,\bket{B^-(p)}
\;=\; i p^\mu f_B \,e^{-ipx}\;.
\end{equation}
Unlike decays into electrons or muons, the chirality-flip factor 
$(m_\tau/m_B)^2\approx 0.1$ is not small, 
but the decay is nevertheless suppressed
due to the smallness of $|V_{ub}|$.
 
If we wish to compare the experimental value, Eq.~(\ref{BtaunuExp}), 
against this SM expression 
instead of using it to extract $|V_{ub}|f_B$, we must obtain
the values of $|V_{ub}|$ and $f_B$ from other sources.
The extraction of $|V_{ub}|$ from charmless semileptonic $B$-decays 
($B\rightarrow\pi\ell\nu$ with $\ell=e$ or $\mu$)
is difficult
requiring considerable theoretical input \cite{Amsler:2008zzb,Kowalewski:2008zz,Petrella:2009tz,HFAG}.
The value quoted in the 2008 Review of Particle Properties \cite{Amsler:2008zzb,Kowalewski:2008zz} is
\begin{equation}
|V_{ub}|_{\mathrm{RPP}} \;=\; (\,3.95\pm0.35\,)\times 10^{-3}\;,
\label{VubRPP}
\end{equation}
with the error dominated by theoretical uncertainty.
Note that in using this value as the SM value of $|V_{ub}|$, we are assuming that
new physics will not affect charmless semileptonic $B$-decay.
The value of $f_B$ is obtained from unquenched lattice QCD.
The HPQCD collaboration reports \cite{Gray:2005ad}
\begin{equation}
f_{B} \;=\; 0.216\pm 0.022\;\mathrm{GeV}\;.
\label{fBHPQCD}
\end{equation}
As can be seen, both $|V_{ub}|$ and $f_B$ suffer from uncertainties on the order of 10\%.
Substituting Eqs.~(\ref{VubRPP}) and (\ref{fBHPQCD}) into Eq.~(\ref{SMprediction}), we find
\begin{equation}
\mathcal{B}(B\rightarrow\tau\nu)_{\mathrm{SM}} \;=\; (\,1.29\pm \,0.35)\times 10^{-4}\;.
\label{SMvalue}
\end{equation}
The UTfit \cite{UTfit} and CKMfit \cite{CKMfit} collaborations constrain this
branching fraction with global SM fits and respectively find 
%
%
%
%
%
\begin{equation}
\begin{array}{ll}
\mathcal{B}(B\rightarrow\tau\nu)_\mathrm{UTfit}  &\;=\; (\,0.81\pm \,0.12)\times 10^{-4}\;,\\
\mathcal{B}(B\rightarrow\tau\nu)_\mathrm{CKMfit} &\;=\; (\,0.92\;{}^{+0.10}_{-0.11})\times 10^{-4}\;, 
\end{array}
\end{equation}
as the SM value.
Though the errors are much smaller, and the central values in disagreement with the
experimental value by almost $2\sigma$, 
we use neither of these values and adhere to Eq.~(\ref{SMvalue}) since
the presence of new physics that would shift $B\rightarrow \tau\nu$ away from the
SM may affect other observables used in the global fits as well.

\subsection{Constraint on New Physics}

The fractional errors on both the experimental value, Eq.~(\ref{BtaunuExp}), and the
SM prediction, Eq.~(\ref{SMvalue}), are large. However, the constraint on new physics
is not necessarily weak since the SM amplitude itself is suppressed by 
$V_{ub}$ requiring new physics effects to be equally suppressed.
We follow Dobrescu and Kronfeld \cite{Dobrescu:2008er} and assume that new physics
effects can be expressed in a model independent way with the effective Lagrangian
\begin{equation}
\mathcal{L}_\mathrm{new} \;=\;
\dfrac{C_A}{M^2}
\bigl(\,\overline{u}\gamma^\mu\gamma_5 b\,\bigr)
\bigl(\,\overline{\tau}_L\gamma_\mu \nu_L\,\bigr)
+
\dfrac{C_P}{M^2}
\bigl(\,\overline{u}\gamma_5 b\,\bigr)
\bigl(\,\overline{\tau}_R \nu_L\,\bigr)
+ h.c.
\label{DobrescuKronfeld}
\end{equation}
where $M$ is the scale of new physics, and 
$C_A$ and $C_P$ are constants that may be complex in general.
Only these operators will cause the decay amplitude from new physics 
to interfere with that from the SM shifting $\sqrt{2}\,G_F V_{ub} m_\tau$ in Eq.~(\ref{SMprediction}) to
\begin{equation}
\sqrt{2}\,G_F V_{ub} m_\tau
\;\rightarrow\;
\sqrt{2}\,G_F V_{ub} m_\tau
+ \dfrac{1}{M^2}
\left( C_A m_\tau - \dfrac{C_P m_B^2}{m_b} \right)
\;,
\end{equation}
where the $u$ quark mass has been neglected.
Assuming that there is no correlation between the experimental and SM values,
Eqs.~(\ref{BtaunuExp}) and (\ref{SMvalue}), we find
\begin{equation}
\dfrac{\mathcal{B}(B\rightarrow\tau\nu)_\mathrm{exp}}
      {\mathcal{B}(B\rightarrow\tau\nu)_\mathrm{SM}}
\;=\; 1.17\pm 0.41\;,
\end{equation}
which translates to
\begin{equation}
\left|\; 1 + \dfrac{1}{\sqrt{2}\,G_F V_{ub} M^2}
\left(C_A - C_P\dfrac{m_B^2}{m_b m_\tau}\right)
\right|^2
\;=\; 1.17\pm 0.41\;.
\end{equation}
%
%
In the standard CKM parametrization 
we have $V_{ub} = |V_{ub}| e^{-i\delta}$, where $\delta$ is the CP 
violating phase \cite{Ceccucci:2008zz}.
Therefore,
\begin{eqnarray}
\left|\; 1 + \dfrac{1}{\sqrt{2}\,G_F V_{ub} M^2}
\left(C_A - C_P\dfrac{m_B^2}{m_b m_\tau}\right)
\right|^2
& \approx &
1 + 2\;
\dfrac{\mathrm{Re}\left[ e^{i\delta}\!\left(C_A - C_P m_B^2/m_b m_\tau\right)\right]}
      {\sqrt{2}\,G_F |V_{ub}|M^2}\;.
\end{eqnarray}
Setting
\begin{equation}
C \;\equiv\;
\mathrm{Re}\left[ e^{i\delta}\!\left(C_A - C_P m_B^2/m_b m_\tau\right)\right]\;,
\label{Cdef}
\end{equation}
the above bound becomes
\begin{equation}
\dfrac{C}{\sqrt{2}\,G_F |V_{ub}|M^2} 
\;=\; \Bigl[(1.53\pm 0.14)\times 10^{3}\Bigr]\,C \left(\dfrac{100\,\mathrm{GeV}}{M}\right)^2
\;=\; 0.09\pm 0.20\;,
\end{equation}
or
\begin{equation}
C \left(\dfrac{100\,\mathrm{GeV}}{M}\right)^2
\;=\; 0.00006\pm 0.00013\;.
\end{equation}
%
%
%
%
%
The 2$\sigma$ (95\%) range of this ratio is therefore 
\begin{equation}
-0.00020
\;<\; 
C \left(\dfrac{100\,\mathrm{GeV}}{M}\right)^2
\;<\; 
0.00032\;,
\qquad
(\mbox{95\% C.L.})\;.
\label{BtaunuBound}
\end{equation}
The bound on the scale of new physics $M$ will
depend on the sign of $C$:
\begin{equation}
\begin{array}{lll}
M/\sqrt{+C} & \ge\; 6\;\mathrm{TeV}\qquad &\mbox{if $C>0$}\;,\\
M/\sqrt{-C} & \ge\; 7\;\mathrm{TeV}\qquad &\mbox{if $C<0$}\;, 
\end{array}\qquad
(\mbox{95\% C.L.})\;.
\label{BtaunuMassBound}
\end{equation}

%
%
%
%


\subsection{Constraints on R-parity violation}

\begin{figure}[ht]
\begin{center}
\begin{picture}(400,120)(-100,-70)
\SetScale{1}
\SetWidth{1}
\ArrowLine(-60,30)(0,30)
\ArrowLine(60,30)(0,30)
\ArrowLine(0,-30)(60,-30)
\ArrowLine(0,-30)(-60,-30)
\DashArrowLine(0,30)(0,-30){5}
\Vertex(0,30){2}
\Vertex(0,-30){2}
\Text(-70,30)[]{$b_L$}
\Text(70,30)[]{$\overline{\nu_{\tau\!L}}$}
\Text(-70,-30)[]{$\overline{u_L}$}
\Text(70,-30)[]{$\tau_L^-$}
\Text(12,0)[]{$\tilde{d}_{kR}$}
\Text(0,40)[]{$\lambda'_{33k}$}
\Text(0,-40)[]{$-\lambda^{\prime *}_{31k}$}
\Text(0,-65)[]{(a)}
\SetOffset(200,0)
\ArrowLine(-60,30)(-30,0)
\ArrowLine(60,30)(30,0)
\ArrowLine(30,0)(60,-30)
\ArrowLine(-30,0)(-60,-30)
\DashArrowLine(-30,0)(30,0){5}
\Vertex(30,0){2}
\Vertex(-30,0){2}
\Text(-70,30)[]{$b_R$}
\Text(70,30)[]{$\overline{\nu_{\tau\!L}}$}
\Text(-70,-30)[]{$\overline{u_L}$}
\Text(70,-30)[]{$\tau_R^-$}
\Text(0,12)[]{$\tilde{e}_{iL}$}
\Text(50,0)[]{$-\lambda_{i33}$}
\Text(-50,0)[]{$-\lambda^{\prime *}_{i13}$}
\Text(0,-65)[]{(b)}
\end{picture}
\end{center}
\caption{Possible R-parity violating contributions to $B^-\rightarrow \tau^-\bar{\nu}_\tau$.
The index is
$k=1$, $2$, or $3$ in (a), while $i=1$ or $2$ in (b) due to the anti-symmetry of $\lambda_{ijk}$
in the first two indices.}
\label{RPV-Bdecay}
\end{figure}
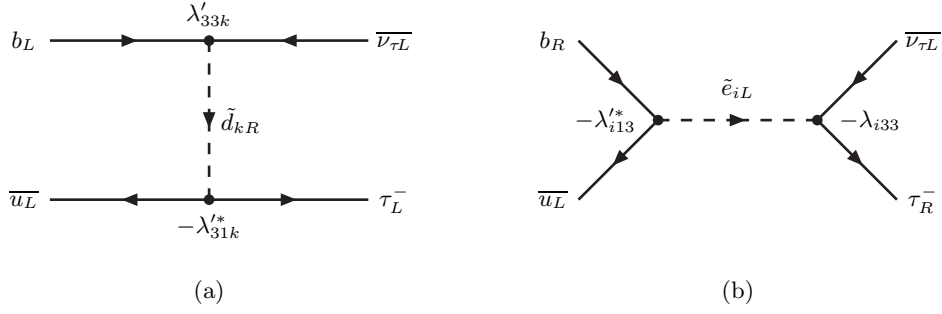

The new physics to which the above bounds apply must distinguish among fermion flavors
since it must affect $B\rightarrow\tau\nu_\tau$ without affecting
$B\rightarrow\pi\ell\nu_\ell$ ($\ell=e$ or $\mu$).
As an example, we consider R-parity violating supersymmetry (SUSY),
the superpotential of which is given by \cite{RPV,Barbier:2004ez}
\begin{equation}
W_{\not R} =\frac{1}{2}\lambda_{ijk}\hat L_i \hat L_j \hat E_k
+\lambda^{\prime}_{ijk}\hat L_i \hat Q_j \hat D_k
+\frac{1}{2}\lambda^{\prime\prime}_{ijk}\hat U_i \hat D_j \hat D_k\;.
\end{equation}
Here $i$, $j$, $k$ are generation indices, while
$SU(2)$-weak isospin and $SU(3)$-color indices are suppressed. 
The coefficients $\lambda_{ijk}$ are antisymmetric in the first two indices, while
$\lambda''_{ijk}$ are antisymmetric in the latter two.
Consequently, there are 9 independent $LLE$ couplings, 
27 independent $LQD$ couplings, and 9 independent $UDD$ couplings.
The decay $B\rightarrow\pi\ell\nu_\ell$ ($\ell=e$ or $\mu$) can be affected by the
coupling combinations 
$\lambda^\prime_{i1k}\lambda^{\prime *}_{i3k}$ ($i=1$ or $2$, $k$ arbitrary) and 
$\lambda_{ijj}\lambda^{\prime *}_{i13}$ ($j=1$ or $2$, $i=3$ or $3-j$)
so these are assumed to be sufficiently small.
The coupling combinations which affect
$B\rightarrow \tau\nu_\tau$ are shown in Figure~\ref{RPV-Bdecay}.
The decay can proceed either via $t$-channel sdown exchange, or via
$s$-channel selectron exchange.

\subsubsection{$t$-channel sdown exchange}

$t$-channel exchange of $\tilde{d}_{kR}$ ($k=1,2$, or $3$) is described by the effective operator
\begin{equation}
\mathcal{L}_{\tilde{d}_{kR}} \;=\;
-\dfrac{\lambda'_{33k}\lambda^{\prime *}_{31k}}{M^2_{\tilde{d}_{kR}}}
\left(\overline{u_{L}^{\phantom{c}}} \tau_{L}^c \right)
\left(\overline{\nu_{\tau\!L}^c} b_{L}^{\phantom{c}} \right) 
\;.
\label{sdownExchange}
\end{equation}
A Fierz transformation allows us to rewrite
\begin{equation}
\left(\overline{u_{L}^{\phantom{c}}} \tau_{L}^c \right)
\left(\overline{\nu_{\tau\!L}^c} b_{L}^{\phantom{c}} \right) 
\;=\;
-\dfrac{1}{2}
\left(\overline{u_{L}^{\phantom{c}}}\gamma_\mu b_{L}^{\phantom{c}} \right) 
\left(\overline{\nu_{\tau\!L}^c}\gamma^\mu \tau_{L}^c \right)
\;=\;
+\dfrac{1}{4}
\bigl(\,\overline{u}\gamma_\mu (1-\gamma_5) b \,\bigr) 
\bigl(\,\overline{\tau}_{L} \gamma^\mu \nu_{\tau L} \,\bigr)
\;.
\end{equation}
The relevant part of the operator Eq.~(\ref{sdownExchange}) is therefore
\begin{equation}
+\dfrac{\lambda'_{33k}\lambda^{\prime *}_{31k}}{4M^2_{\tilde{d}_{kR}}}
\bigl(\,\overline{u}\gamma_\mu \gamma_5 b \,\bigr) 
\bigl(\,\overline{\tau}_{L} \gamma^\mu \nu_{\tau L} \,\bigr)
\;.
\label{sdownExchange2}
\end{equation}
Comparison with Eqs.~(\ref{DobrescuKronfeld}) and (\ref{Cdef}) leads to the identifications
\begin{equation}
M\;=\;M_{\tilde{d}_{kR}}\;,\qquad
C_A \;=\; \dfrac{\lambda'_{33k}\lambda^{\prime *}_{31k}}{4}\;,\qquad
C \;=\; \mathrm{Re}\left[ e^{i\delta} C_A \right]
\;=\;
\dfrac{\mathrm{Re}
\left[ e^{i\delta}\lambda'_{33k}\lambda^{\prime *}_{31k} \right]}{4}\;.
\end{equation}
Allowing only one sdown contribution to be non-zero at a time,
the bounds of Eq.~(\ref{BtaunuBound}) translate to
\begin{equation}
-0.0008\;
\;<\;
\mathrm{Re}\left[ e^{i\delta}\lambda'_{33k}\lambda^{\prime *}_{31k} \right]
\left(\dfrac{100\,\mathrm{GeV}}{M_{\tilde{d}_{kR}}}\right)^2
\;<\; 
+0.0013\;,
\qquad
(\mbox{95\% C.L.})\;.
\end{equation}
The bounds on the sdown mass are
\begin{equation}
\begin{array}{lll}
\dfrac{M_{\tilde{d}_{kR}}}{\sqrt{+\mathrm{Re}
\left[ e^{i\delta}\lambda'_{33k}\lambda^{\prime *}_{31k} \right]}} & \ge\; 3\;\mathrm{TeV}\qquad &\mbox{if $\mathrm{Re}
\left[ e^{i\delta}\lambda'_{33k}\lambda^{\prime *}_{31k} \right]>0$}\;,\\
\dfrac{M_{\tilde{d}_{kR}}}{\sqrt{-\mathrm{Re}
\left[ e^{i\delta}\lambda'_{33k}\lambda^{\prime *}_{31k} \right]}} & \ge\; 4\;\mathrm{TeV}\qquad &\mbox{if $\mathrm{Re}
\left[ e^{i\delta}\lambda'_{33k}\lambda^{\prime *}_{31k} \right]<0$}\;,
\end{array}
\qquad
(\mbox{95\% C.L.})\;.
\label{BtaunuRPVbound}
\end{equation}
\subsubsection{$s$-channel selectron exchange}

$s$-channel exchange of $\tilde{e}_{iL}$ ($i=1$ or $2$) is described by the effective operator
\begin{equation}
\mathcal{L}_{\tilde{e}_{iL}} \;=\;
\dfrac{\lambda_{i33}^{\phantom{*}}\lambda^{\prime *}_{i13}}{M^2_{\tilde{e}_{iL}}}
\bigl(\,\overline{u}_{L} b_{R} \,\bigr) 
\bigl(\,\overline{\tau}_{R}\nu_{\tau L} \,\bigr)
\;=\;
\dfrac{\lambda_{i33}^{\phantom{*}}\lambda^{\prime *}_{i13}}{2M^2_{\tilde{e}_{iL}}}
\bigl(\,\overline{u}(1+\gamma_5) b\,\bigr) 
\bigl(\,\overline{\tau}_{R}\nu_{\tau L} \,\bigr)
\;,
\label{selectronExchange}
\end{equation}
the relevant part of which is
\begin{equation}
\dfrac{\lambda_{i33}^{\phantom{*}}\lambda^{\prime *}_{i13}}{2M^2_{\tilde{e}_{iL}}}
\bigl(\,\overline{u}\gamma_5 b\,\bigr) 
\bigl(\,\overline{\tau}_{R}\nu_{\tau L} \,\bigr)
\;.
\label{selectronExchange2}
\end{equation}
Comparison with Eqs.~(\ref{DobrescuKronfeld}) and (\ref{Cdef}) leads to the identifications
\begin{equation}
M\;=\;M_{\tilde{e}_{iL}}\;,\qquad
C_P \;=\; \dfrac{\lambda_{i33}^{\phantom{*}}\lambda^{\prime *}_{i13}}{2}\;,\qquad
C \;=\; \dfrac{m_B^2}{m_b m_\tau}\mathrm{Re}\left[ -e^{i\delta} C_P \right]
\;=\;
\dfrac{m_B^2}{2 m_b m_\tau}
\mathrm{Re}\left[ -e^{i\delta}\lambda_{i33}^{\phantom{*}}\lambda^{\prime *}_{i13} \right]
\;.
\end{equation}
The factor $m_B^2/m_b m_\tau$ is equal to \cite{Manohar:2008zz}
\begin{equation}
\dfrac{m_B^2}{m_b m_\tau}
\;=\; 
\dfrac{(\,5.27917\pm 0.00029\;\mathrm{GeV}\,)^2}
      {(\,1.77684\pm 0.00017\;\mathrm{GeV}\,)(\,4.79\,{}^{+0.19}_{-0.08}\,\mathrm{GeV}\,)}
\;=\; 3.27\,{}^{+0.06}_{-0.12}\;,
\end{equation}
where we have used the pole mass for $m_b$.
Allowing only one selectron contribution to be non-zero at a time,
the bounds of Eq.~(\ref{BtaunuBound}) translate to
\begin{equation}
-0.00012\;
\;<\;
\mathrm{Re}\left[ -e^{i\delta}\lambda_{i33}^{\phantom{*}}\lambda^{\prime *}_{i13} \right]
\left(\dfrac{100\,\mathrm{GeV}}{M_{\tilde{e}_{iL}}}\right)^2
\;<\; 
+0.00020\;
\;,
\qquad
(\mbox{95\% C.L.})\;.
\end{equation}
The corresponding bounds on the selectron mass are
\begin{equation}
\begin{array}{lll}
\dfrac{M_{\tilde{e}_{iL}}}{\sqrt{+\mathrm{Re}
\left[ -e^{i\delta}\lambda_{i33}^{\phantom{*}}\lambda^{\prime *}_{i13} \right]}} & \ge\; 7\;\mathrm{TeV}\qquad &\mbox{if $\mathrm{Re}
\left[ -e^{i\delta}\lambda_{i33}^{\phantom{*}}\lambda^{\prime *}_{i13} \right]>0$}\;,\\
\dfrac{M_{\tilde{e}_{iL}}}{\sqrt{-\mathrm{Re}
\left[ -e^{i\delta}\lambda_{i33}^{\phantom{*}}\lambda^{\prime *}_{i13} \right]}} & \ge\; 9\;\mathrm{TeV}\qquad &\mbox{if $\mathrm{Re}
\left[ -e^{i\delta}\lambda_{i33}^{\phantom{*}}\lambda^{\prime *}_{i13} \right]<0$}\;,
\end{array}
\qquad
(\mbox{95\% C.L.})\;.
\label{BtaunuRPVbound2}
\end{equation}
%

\section{$\tau^-\rightarrow \ell^- K^0_S$}

\subsection{Experimental Bounds on Lepton Flavor Violating $\bm{\tau}$ Decays}

Babar recently reported their measurements for tau lepton-flavor-violating decays $\tau^- \rightarrow \ell^- K_S^0$ ($\ell = e$ or $\mu$)  using a data sample corresponding to an integrated luminosity of $469\,\mathrm{fb}^{-1}$. The upper limits on the branching fractions for the two channels, at 90\% confidence level, are \cite{Aubert:2009ys}
\begin{equation}
\begin{array}{ll}
\mathcal{B}(\tau^- \rightarrow \mu^- K_S^0)_\mathrm{Babar} & < \; 4.0 \times 10^{-8}\;, \\
\mathcal{B}(\tau^- \rightarrow e^-   K_S^0)_\mathrm{Babar} & < \; 3.3 \times 10^{-8}\;,
\end{array}
\qquad
(\mbox{90\% C.L.})\;.
\label{BabarLFV90bound}
\end{equation}
These supercede the previous 90\% bounds from Belle based on $281\,\mathrm{fb}^{-1}$ of data, which were 
\cite{Miyazaki:2006sx}
\begin{equation}
\begin{array}{ll}
\mathcal{B}(\tau^- \rightarrow \mu^- K_S^0)_\mathrm{Belle} & < \; 4.9 \times 10^{-8}\;, \\
\mathcal{B}(\tau^- \rightarrow e^-   K_S^0)_\mathrm{Belle} & < \; 5.6 \times 10^{-8}\;,
\end{array}
\qquad
(\mbox{90\% C.L.})\;.
\end{equation}
For the sake of comparison with the bounds from $B\rightarrow\tau\nu$ we derived in the previous section, 
and also with the previous bounds from Ref.~\cite{Dreiner:2006gu}, we will use the 95\% confidence level bounds  
from Babar, which can be read off from Fig.~4 of Ref.~\cite{Aubert:2009ys} as
\begin{equation}
\begin{array}{ll}
\mathcal{B}(\tau^- \rightarrow \mu^- K_S^0)_\mathrm{Babar} & < \; 5.2 \times 10^{-8}\;, \\
\mathcal{B}(\tau^- \rightarrow e^-   K_S^0)_\mathrm{Babar} & < \; 4.3 \times 10^{-8}\;,
\end{array}
\qquad
(\mbox{95\% C.L.})\;.
\label{BabarLFVbound}
\end{equation}
Since there exist no SM contribution to these processes, these bounds translate directly 
into bounds on new physics.

\subsection{Constraints on R-parity violation}

\begin{figure}[ht]
\begin{center}
\begin{picture}(600,110)(-70,-60)
\SetScale{1}
\SetWidth{1}
\ArrowLine(-50,30)(0,30)
\ArrowLine(0,30)(50,30)
\ArrowLine(50,10)(0,-10)
\ArrowLine(0,-10)(50,-30)
\DashArrowLine(0,-10)(0,30){5}
\Vertex(0,30){2}
\Vertex(0,-10){2}
\Text(-60,30)[]{$\tau^-_L$}
\Text(60,30)[]{$d_{kR}$}
\Text(60,10)[]{$\overline{d_{\ell R}}$}
\Text(60,-30)[]{$e_{jL}^-$}
\Text(12,10)[]{$\tilde{u}_{iL}$}
\Text(0,40)[]{$-\lambda'_{3ik}$}
\Text(0,-20)[]{$-\lambda^{\prime *}_{ji\ell}$}
\Text(0,-55)[]{(a)}
\SetOffset(180,0)
\ArrowLine(-50,30)(0,30)
\ArrowLine(0,30)(50,30)
\ArrowLine(50,10)(0,-10)
\ArrowLine(0,-10)(50,-30)
\DashArrowLine(0,-10)(0,30){5}
\Vertex(0,30){2}
\Vertex(0,-10){2}
\Text(-60,30)[]{$\tau^-_L$}
\Text(60,30)[]{$e_{jR}^-$}
\Text(60,10)[]{$\overline{d_{\ell R}}$}
\Text(60,-30)[]{$d_{kL}$}
\Text(12,10)[]{$\tilde{\nu}_{iL}$}
\Text(0,40)[]{$\lambda_{i3j}$}
\Text(0,-20)[]{$\lambda^{\prime *}_{ik\ell}$}
\Text(0,-55)[]{(b)}
\SetOffset(360,0)
\ArrowLine(-50,30)(0,30)
\ArrowLine(0,30)(50,30)
\ArrowLine(50,10)(0,-10)
\ArrowLine(0,-10)(50,-30)
\DashArrowLine(0,30)(0,-10){5}
\Vertex(0,30){2}
\Vertex(0,-10){2}
\Text(-60,30)[]{$\tau^-_R$}
\Text(60,30)[]{$e_{jL}^-$}
\Text(60,10)[]{$\overline{d_{k L}}$}
\Text(60,-30)[]{$d_{\ell R}$}
\Text(12,10)[]{$\tilde{\nu}_{iL}$}
\Text(0,40)[]{$\lambda^*_{ij3}$}
\Text(0,-20)[]{$\lambda'_{ik\ell}$}
\Text(0,-55)[]{(c)}
\end{picture}
\end{center}
\caption{Possible R-parity violating contributions to $\tau^-\rightarrow \ell^- K^0_S$,
($\ell=e$ or $\mu$).
The indices are $j=1$ or 2, $(k\ell)=(12)$ or $(21)$.
The index $i$ for the sup exchange diagram can take on any value from 1 to 3,
but that in the sneutrino exchange diagrams is restricted
due to the antisymmetry of $\lambda_{ijk}$ in the
first two indices and only two values are possible for each diagram:
$i=1$ or $2$ in (b), and $i=3$ or $3-j$ in (c).
}
\label{RPV-taudecay}
\end{figure}
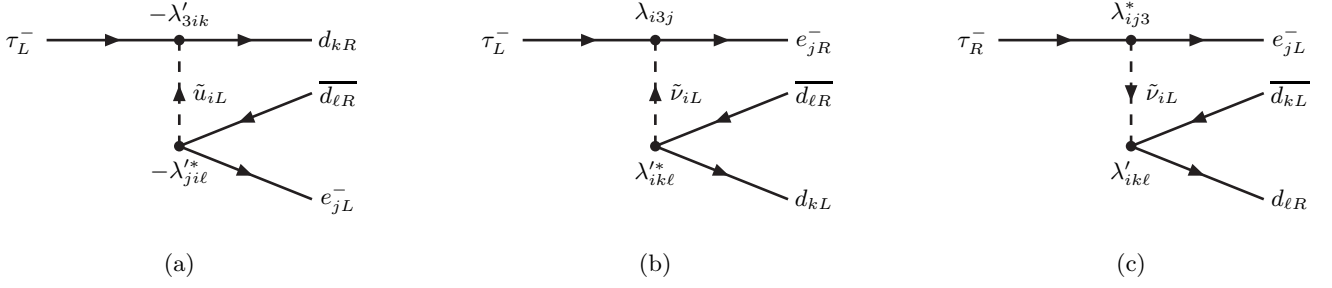

As in the $B\rightarrow \tau\nu_\tau$ analysis, we use R-parity violating SUSY as an example.
Possible contributions to the process $\tau^-\rightarrow \ell^- K^0_S$
from R-parity violation are shown in Figure~\ref{RPV-taudecay}.
The decay can proceed either via sup exchange, or sneutrino exchange.

\subsubsection{sup exchange}

For definiteness, let us first consider the decay $\tau^-\rightarrow \mu^- K^0_S$ via the 
sup exchange subprocess
$\tau_L^-\rightarrow \mu_L^- \overline{s_R^{\phantom{*}}}d_R^{\phantom{*}}$ or
$\tau_L^-\rightarrow \mu_L^- \overline{d_R^{\phantom{*}}}s_R^{\phantom{*}}$
shown in Figure~\ref{RPV-taudecay}(a).
The indices for these subprocesses are $j=2$ with $(k\ell)=(12)$ or $(21)$.
The effective operator induced by $\tilde{u}_{iL}$ ($i=1,2$, or $3$) exchange is
\begin{equation}
\mathcal{L}_{\tilde{u}_{iL}}
\;=\;
\dfrac{\lambda'_{3ik}\lambda^{\prime *}_{2i\ell}}{M^2_{\tilde{u}_{iL}}}
\bigl(\,\overline{d_{kR}} \tau_{L}\,\bigr)
\bigl(\,\overline{\mu_{L}} d_{\ell R}\,\bigr)
\;.
\end{equation}
A Fierz transformation allows us to rewrite
\begin{equation}
\bigl(\,\overline{d_{kR}} \tau_{L}\,\bigr)
\bigl(\,\overline{\mu_{L}} d_{\ell R}\,\bigr)
\;=\;
\dfrac{1}{2}
\bigl(\,\overline{d_{kR}}\gamma_\mu d_{\ell R}\,\bigr)
\bigl(\,\overline{\mu_{L}}\gamma^\mu \tau_{L}\,\bigr)
\;=\;
\dfrac{1}{4}
\bigl(\,\overline{d_k}\gamma_\mu(1+\gamma_5) d_\ell\,\bigr)
\bigl(\,\overline{\mu_{L}}\gamma^\mu \tau_{L}\,\bigr)
\;,
\end{equation}
and the part of the operator relevant for the decay in question is
\begin{equation}
\dfrac{\lambda'_{3ik}\lambda^{\prime *}_{2i\ell}}{4M^2_{\tilde{u}_{iL}}}
\bigl(\,\overline{d_k}\gamma_\mu\gamma_5 d_\ell\,\bigr)
\bigl(\,\overline{\mu_{L}}\gamma^\mu \tau_{L}\,\bigr)
\;.
\label{supExchange}
\end{equation}
The matrix element of $\overline{d_\ell}\gamma^\mu\gamma_5 d_k$ between the vacuum and the
$K^0_S = (K^0+\overline{K^0})/\sqrt{2}$ state can be expressed as
\begin{equation}
\bbra{K^0_S(p)}\,\overline{d_\ell}(x)\gamma^\mu\gamma_5 d_k(x)\,\bket{0}
\;=\; -\dfrac{i}{\sqrt{2}}\, p^\mu f_{K^0} \,e^{ipx}\;,
\end{equation}
where the $K^0$ decay constant $f_{K^0}$ is defined as 
\begin{equation}
\bbra{0}\,\overline{s}(x)\gamma^\mu\gamma_5 d(x)\,\bket{K^0(p)} \;=\;
\bbra{0}\,\overline{d}(x)\gamma^\mu\gamma_5 s(x)\,\bket{\overline{K^0}(p)} \;=\; 
i p^\mu f_{K^0} \,e^{-ipx}\;.
\end{equation}
The $\tau^-\rightarrow \mu^- K^0_S$ branching fraction due to the operator
Eq.~(\ref{supExchange}) is then expressed as \cite{Dreiner:2006gu}
\begin{equation}
\mathcal{B}(\tau^-\rightarrow \mu^- K_{S}^0)
\;=\; \dfrac{\left|\lambda'_{3ik}\lambda^{\prime *}_{2i\ell}\right|^2}{M^4_{\tilde{u}_{iL}}}\;
\dfrac{\sqrt{\Lambda(m_\tau^2,m_\mu^2,m_K^2)} \Bigl[\,(m_\tau^2-m_\mu^2)^2-m_K^2(m_\tau^2+m_\mu^2)\,\Bigr]}
{1024\pi\, m_\tau^3}\; (f_{K^0})^2\,\tau_\tau\;,
\end{equation}
where
\begin{equation}
\Lambda(a,b,c)\;\equiv\;
a^2 + b^2 + c^2 - 2ab - 2bc - 2ca\;.
\end{equation}
Invoking isospin symmetry, we assume that $f_{K^0}$ is equal to the decay constant of the
charged Kaons $f_{K^\pm}=0.1555\pm 0.0008\,\mathrm{GeV}$ \cite{Rosner:2008yu} and find
\begin{equation}
\mathcal{B}(\tau^-\rightarrow \mu^- K_{S}^0)
\;=\; (\,0.1561\pm 0.0017\,)\,\left|\lambda'_{3ik}\lambda^{\prime *}_{2i\ell}\right|^2
\left(\dfrac{100\,\mathrm{GeV}}{M_{\tilde{u}_{iL}}}\right)^4\;.
\end{equation}
Then, the 95\% Babar bound, Eq.~(\ref{BabarLFVbound}), translates to
\begin{equation}
\sqrt{\left|\lambda'_{3ik}\lambda^{\prime *}_{2i\ell}\right|}
\left(\dfrac{100\,\mathrm{GeV}}{M_{\tilde{u}_{iL}}}\right)
\;<\; 0.024\;,
\qquad
\mbox{or}
\qquad
\dfrac{M_{\tilde{u}_{iL}}}{\sqrt{\left|\lambda'_{3ik}\lambda^{\prime *}_{2i\ell}\right|}}
\;>\; 4.2\,\mathrm{TeV}\;,
\qquad
(\mbox{95\% C.L.})
\;.
\end{equation}
Similarly, the branching fraction of
$\tau^-\rightarrow e^- K^0_S$ proceeding via the subprocesses 
$\tau_L^-\rightarrow e_L^- \overline{s_R^{\phantom{*}}}d_R^{\phantom{*}}$ and
$\tau_L^-\rightarrow e_L^- \overline{d_R^{\phantom{*}}}s_R^{\phantom{*}}$ 
is given by
\begin{eqnarray}
\mathcal{B}(\tau^-\rightarrow e^- K_{S}^0)
& = &
\dfrac{\left|\lambda'_{3ik}\lambda^{\prime *}_{1i\ell}\right|^2}{M^4_{\tilde{u}_{iL}}}\;
\dfrac{\sqrt{\Lambda(m_\tau^2,m_e^2,m_K^2)} \Bigl[\,(m_\tau^2-m_e^2)^2-m_K^2(m_\tau^2+m_e^2)\,\Bigr]}
{1024\pi\, m_\tau^3}\; (f_{K^0})^2\,\tau_\tau \cr
& = & (\,0.1581\pm 0.0017\,)\,\left|\lambda'_{3i1}\lambda^{\prime *}_{1i2}\right|^2
\left(\dfrac{100\,\mathrm{GeV}}{M_{\tilde{u}_{iL}}}\right)^4\;.
\end{eqnarray}
The constraint from the 95\% Babar bound, Eq.~(\ref{BabarLFVbound}), is then
\begin{equation}
\sqrt{\left|\lambda'_{3ik}\lambda^{\prime *}_{1i\ell}\right|}
\left(\dfrac{100\,\mathrm{GeV}}{M_{\tilde{u}_{iL}}}\right)
\;<\; 0.023\;,
\qquad
\mbox{or}
\qquad
\dfrac{M_{\tilde{u}_{iL}}}{\sqrt{\left|\lambda'_{3ik}\lambda^{\prime *}_{1i\ell}\right|}}
\;>\; 4.4\,\mathrm{TeV}\;,
\qquad
(\mbox{95\% C.L.})
\;.
\end{equation}
%

\subsubsection{sneutrino exchange}

Next, we consider the decay $\tau^-\rightarrow \mu^- K^0_S$ via the 
sneutrino exchange subprocess
$\tau_L^-\rightarrow \mu_R^- \overline{s_R^{\phantom{*}}}d_L^{\phantom{*}}$ or
$\tau_L^-\rightarrow \mu_R^- \overline{d_R^{\phantom{*}}}s_L^{\phantom{*}}$ 
shown in Figure~\ref{RPV-taudecay}(b).
The indices for these subprocesses are $j=2$ with $(k\ell)=(12)$ or $(21)$.
The effective operator induced by $\tilde{\nu}_{iL}$ ($i=1$ or $2$) exchange is
\begin{equation}
\mathcal{L}_{\tilde{\nu}_{iL}}
\;=\; \dfrac{\lambda_{i32}^{\phantom{*}}\lambda^{\prime *}_{ik\ell}}{M^2_{\tilde{\nu}_{iL}}}
\bigl(\,\overline{\mu_{R}} \tau_{L}\,\bigr)
\bigl(\,\overline{d_{kL}} d_{\ell R}\,\bigr)
\;=\; \dfrac{\lambda_{i32}^{\phantom{*}}\lambda^{\prime *}_{ik\ell}}{2M^2_{\tilde{\nu}_{iL}}}
\bigl(\,\overline{\mu_{R}} \tau_{L}\,\bigr)
\bigl(\,\overline{d_{k}}(1+\gamma_5)d_{\ell}\,\bigr)\;.
\end{equation}
The part of this operator that is relevant for the decay is 
\begin{equation}
\dfrac{\lambda_{i3j}^{\phantom{*}}\lambda^{\prime *}_{ik\ell}}{2M^2_{\tilde{\nu}_{iL}}}
\bigl(\,\overline{\mu_{R}} \tau_{L}\,\bigr)
\bigl(\,\overline{d_{k}}\gamma_5 d_{\ell}\,\bigr)\;,
\end{equation}
leading to the branching fraction \cite{Dreiner:2006gu}
\begin{equation}
\mathcal{B}(\tau^-\rightarrow \mu^- K_{S}^0)
\;=\;
\dfrac{\left|\lambda_{i32}^{\phantom{*}}\lambda^{\prime *}_{ik\ell}\right|^2}{M^4_{\tilde{\nu}_{iL}}}\;
\dfrac{\sqrt{\Lambda(m_\tau^2,m_\mu^2,m_K^2)} \Bigl(\,m_\tau^2+m_\mu^2-m_K^2\,\Bigr)m_K^2}
{256\pi\, m_\tau^3}
\;\xi^2 (f_{K^0})^2\,\tau_\tau \;,
\label{BtaumuK}
\end{equation}
where the factor $\xi$ is defined as:
\begin{equation}
\xi\;\equiv\;
\dfrac{m_K}{m_d+m_s}
\;\approx\; \dfrac{m_K}{m_s}
\;=\; \dfrac{(496.614\pm 0.024\,\mathrm{MeV})}{(105\,{}^{+25}_{-35}\,\mathrm{MeV})}
\;=\; 4\sim 7\;.
\end{equation}
Here, we have used the $\overline{\mathrm{MS}}$ mass at $\mu=2\,\mathrm{GeV}$ for $m_s$.
The error introduced by the neglect of $m_d$ is only about 5\%.
Allowing $\xi$ to sweep this range, we find
\begin{equation}
\mathcal{B}(\tau^-\rightarrow \mu^- K_{S}^0)
\;=\; \bigl(\,0.8\sim 2.4\,\bigr)
\left|\lambda_{i32}^{\phantom{*}}\lambda^{\prime *}_{ik\ell}\right|^2
\left(\dfrac{100\,\mathrm{GeV}}{M_{\tilde{\nu}_{iL}}}\right)^4\;,
\end{equation}
and Eq.~(\ref{BabarLFVbound}) translates to
\begin{equation}
\sqrt{\left|\lambda_{i32}^{\phantom{*}}\lambda^{\prime *}_{ik\ell}\right|}
\left(\dfrac{100\,\mathrm{GeV}}{M_{\tilde{\nu}_{iL}}}\right)
\;<\; 0.012\sim 0.016\;,\qquad
\dfrac{M_{\tilde{\nu}_{iL}}}
{\sqrt{\left|\lambda_{i32}^{\phantom{*}}\lambda^{\prime *}_{ik\ell}\right|}}
\;>\; (6\sim 8)\,\mathrm{TeV}\;,\qquad
(\mbox{95\% C.L.})
\;.
\end{equation}
The branching fraction due to the subprocesses
$\tau_R^-\rightarrow \mu_L^- \overline{s_L^{\phantom{*}}}d_R^{\phantom{*}}$ or
$\tau_R^-\rightarrow \mu_L^- \overline{d_L^{\phantom{*}}}s_R^{\phantom{*}}$ 
shown in Figure~\ref{RPV-taudecay}(c) is the same as Eq.~(\ref{BtaumuK}) except with the
coupling constants replaced by the combination $\lambda_{i23}^{\phantom{*}}\lambda^{\prime *}_{ik\ell}$
with $i=1$ or $3$, to which the exact same bounds apply.

The analysis for the decay $\tau^-\rightarrow e^- K^0_S$ proceeds in an exactly analogous fashion
and the results are
\begin{equation}
\sqrt{\left|\lambda_{i31}^{\phantom{*}}\lambda^{\prime *}_{ik\ell}\right|}
\left(\dfrac{100\,\mathrm{GeV}}{M_{\tilde{\nu}_{iL}}}\right)
\;<\; 0.011\sim 0.015\;,\qquad
\dfrac{M_{\tilde{\nu}_{iL}}}
{\sqrt{\left|\lambda_{i31}^{\phantom{*}}\lambda^{\prime *}_{ik\ell}\right|}}
\;>\; (7\sim 9)\,\mathrm{TeV}\;,\qquad
(\mbox{95\% C.L.})
\;,
\end{equation}
with $i=1$ or $2$. The same bounds apply to 
$\lambda_{i13}^{\phantom{*}}\lambda^{\prime *}_{ik\ell}$ with $i=2$ or $3$.

\section{Summary \& Discussion}

In Table~\ref{BoundComparison} we list the bounds on 
various R-parity violating coupling combinations 
obtained in this work against those 
obtained by Dreiner et al. in 2002 \cite{Dreiner:2001kc} and in 2006 \cite{Dreiner:2006gu}.
All sfermion masses have been set to 100~GeV.
The bounds from both $B\rightarrow\tau\nu_\tau$ and $\tau\rightarrow\ell K^0_S$ have all improved by factors of $4\sim 5$.
The corresponding lower bound on the scale of new physics is in the 4 to 10 TeV range if we set
all coupling constants to one.

The current single-coupling bounds on the individual 
R-parity violating couplings that appear in Table~\ref{BoundComparison}
are listed in Table~\ref{SingleCouplingBounds}.
The numbers have been updated from those in Table 6.1 on page 110 of Ref.~\cite{Barbier:2004ez}
using the most recent data \cite{Amsler:2008zzb,Porsev:2009pr,Swagato}.
Detailed derivations will be provided elsewhere \cite{KT2,KT3}.
The products of these single-coupling bounds are listed in the rightmost column of 
Table~\ref{BoundComparison} with all sparticle masses set to 100~GeV.
As can be seen, the bounds from $B\rightarrow\tau\nu_\tau$ and
$\tau\rightarrow \ell K_S^0$ are much stronger than the products of the single-coupling bounds 
with the exception of the combination $|\lambda'_{111}\lambda^{\prime *}_{312}|$
for which $|\lambda'_{111}|$ is strongly constrained by neutrinoless double beta-decay.

The experimental error on $B\rightarrow\tau\nu_\tau$ can be expected to be reduced further as more
Belle and Babar data is analyzed.  However, unless the theoretical uncertainty of its
SM prediction based on charmless semileptonic $B$ decay and lattice calculations 
can be reduced also, any improvement on the new physics bounds will be limited.
The decay $\tau\rightarrow \ell K_S^0$, on the other hand, has no SM counterpart, and
any reduction of its experimental upper bound will translate directly into an improvement of the
bounds on new physics.

\section*{Acknowledgements}

We thank Swagato Banerjee, Marcella Bona, Herbert Dreiner, and Martin Hirsch 
for helpful discussions and communications.
A portion of this work was presented by Kao at Pheno 2009. We thank the organizers for providing us with the
opportunity to present our results.
This work was supported by the U.S. Department of Energy, 
grant DE--FG05--92ER40709, Task A.

\begin{turnpage}

\begingroup

\begin{table}[p]
\begin{tabular}{|c|c|c|c|c|l|}
\hline
$\lambda'\lambda'$ & decay & \ sparticle \ & new bound & \ previous bound \ [Ref] (year)\ & \ Product of single-coupling $2\sigma$ bounds\ \ \\
\hline\hline
\ $(31k)(33k)$ \ & \ $B\rightarrow\tau\nu_\tau$ \ & $\tilde{d}_{kR}$
& \ $-0.8\times 10^{-3} < \mathrm{Re}[\,e^{i\delta}\lambda'\lambda^{\prime *}] < 1.3\times 10^{-3}$ \ & \ N/A \ & \ $0.03$ $[R_{\tau\pi}][R_{\tau}^Z]$ \ \\
\hline
\ $(211)(312)$ \ & \ $\tau^-\rightarrow\mu^- K^0_S$ \ & $\tilde{u}_{1L}$ & $|\lambda'\lambda^{\prime *}| < 5.8\times 10^{-4}$ & $2.4\times 10^{-3}$ \ \cite{Dreiner:2006gu} (2006)\ & \ $0.004$ $[R_{\pi}][R_{\tau\pi}]$ \ \\
\ $(212)(311)$ \ & & $\tilde{u}_{1L}$ & & & \ $0.004$ $[R_{\pi}][R_{\tau\pi}]$ \ \\
\cline{3-3}
\ $(221)(322)$ \ & & $\tilde{u}_{2L}$ & & & \ $0.03$ $[R_{D^0}][R_{D_s}(\tau\mu)]$ \ \\
\ $(222)(321)$ \ & & $\tilde{u}_{2L}$ & & & \ $0.03$ $[R_{D^0}][R_{D_s}(\tau\mu)]$ \ \\
\cline{3-3}
\ $(231)(332)$ \ & & $\tilde{u}_{3L}$ & & & \ $0.3$ $[R_\mu^Z][R_\tau^Z]$ \ \\
\ $(232)(331)$ \ & & $\tilde{u}_{3L}$ & & & \ $0.3$ $[R_\mu^Z][R_\tau^Z]$ \ \\ 
\hline
\ $(111)(312)$ \ & \ $\tau^-\rightarrow e^- K^0_S$ \ & $\tilde{u}_{1L}$ & $|\lambda'\lambda^{\prime *}| < 5.2\times 10^{-4}$ & $2.3\times 10^{-3}$ \ \cite{Dreiner:2006gu} (2006)\ & \ $4\times 10^{-5}$ $[\beta\beta 0\nu][R_{\tau\pi}]$ \ \\
\ $(112)(311)$ \ & & $\tilde{u}_{1L}$ & & & \ $0.002$ $[V_{us},R_\pi][R_{\tau\pi}]$ \ \\
\cline{3-3}
\ $(121)(322)$ \ & & $\tilde{u}_{2L}$ & & & \ $0.01$ $[Q_W({}^{133}\mathrm{Cs})][R_{D_s}(\tau\mu)]$ \ \\
\ $(122)(321)$ \ & & $\tilde{u}_{2L}$ & & & \ $0.06$ $[R_{D^+}][R_{D_s}(\tau\mu)]$ \ \\
\cline{3-3}
\ $(131)(332)$ \ & & $\tilde{u}_{3L}$ & & & \ $0.02$ $[Q_W({}^{133}\mathrm{Cs})][R_\tau^Z]$ \ \\
\ $(132)(331)$ \ & & $\tilde{u}_{3L}$ & & & \ $0.2$ $[A_\mathrm{FB}^s][R_\tau^Z]$ \ \\
\hline\hline
$\lambda\lambda'$ & decay & \ sparticle \ & new bound & \ previous bound \ [Ref] (year)\ & \ Product of single-coupling $2\sigma$ bounds\ \ \\
\hline\hline
\ $(133)(113)$ \ & \ $B\rightarrow\tau\nu_\tau$ \ & $\tilde{e}_{1L}$ & \ $-1.2\times 10^{-4} < \mathrm{Re}[-e^{i\delta}\lambda\lambda^{\prime *}] < 2.0\times 10^{-4}$ \ & 
\ $-6\times 10^{-4} < \mathrm{Re}[\lambda\lambda^{\prime *}] < 1\times 10^{-3}$ \ & \ $0.002$ $[R_\tau][V_{ud},R_\pi]$ \ \\
\ $(233)(213)$ \ & & $\tilde{e}_{2L}$ & & \ \cite{Dreiner:2001kc} (2002)\ & \ $0.003$ $[R_\tau][R_\pi]$ \ \\
\hline
\ $(123)(112)$ \ & \ $\tau^-\rightarrow\mu^- K^0_S$ \ & $\tilde{\nu}_{1L}$ & $|\lambda'\lambda^{\prime *}| < (1.5\sim 2.6)\times 10^{-4}$ & \ $1.0\times 10^{-3}$ \ \cite{Dreiner:2006gu} (2006)\  & \ $0.001$ $[V_{ud}][R_\pi,Q_W({}^{133}\mathrm{Cs})]$ \ \\
\ $(123)(121)$ \ & & $\tilde{\nu}_{1L}$ & & & \ $0.001$ $[V_{ud}][Q_W({}^{133}\mathrm{Cs})]$ \ \\
\ $(132)(112)$ \ & & $\tilde{\nu}_{1L}$ & & & \ $0.002$ $[R_\tau][V_{ud},R_\pi]$ \ \\
\ $(132)(121)$ \ & & $\tilde{\nu}_{1L}$ & & & \ $0.002$ $[R_\tau][Q_W({}^{133}\mathrm{Cs})]$ \ \\
\cline{3-3}
\ $(232)(212)$ \ & & $\tilde{\nu}_{2L}$ & & & \ $0.003$ $[R_\tau][R_\pi]$ \ \\
\ $(232)(221)$ \ & & $\tilde{\nu}_{2L}$ & & & \ $0.005$ $[R_\tau][R_{D^0}]$ \ \\
\cline{3-3}
\ $(233)(312)$ \ & & $\tilde{\nu}_{3L}$ & & & \ $0.003$ $[R_\tau][R_{\tau\pi}]$ \ \\
\ $(233)(321)$ \ & & $\tilde{\nu}_{3L}$ & & & \ $0.02$ $[R_\tau][R_{D_s}(\tau\mu)]$ \\
\hline
\ $(123)(212)$ \ & \ $\tau^-\rightarrow e^- K^0_S$ \ & $\tilde{\nu}_{2L}$ & $|\lambda'\lambda^{\prime *}| < (1.3\sim 2.3)\times 10^{-4}$ & \ $9.7\times 10^{-4}$ \ \cite{Dreiner:2006gu} (2006)\ & \ $0.002$ $[V_{ud}][R_\pi]$ \ \\
\ $(123)(221)$ \ & & $\tilde{\nu}_{2L}$ & & & \ $0.003$ $[V_{ud}][R_{D^0}]$ \\
\cline{3-3}
\ $(131)(112)$ \ & & $\tilde{\nu}_{1L}$ & & & \ $0.002$ $[R_\tau][V_{ud},R_\pi]$ \ \\
\ $(131)(121)$ \ & & $\tilde{\nu}_{1L}$ & & & \ $0.002$ $[R_\tau][Q_W({}^{133}\mathrm{Cs})]$ \ \\
\cline{3-3}
\ $(133)(312)$ \ & & $\tilde{\nu}_{3L}$ & & & \ $0.003$ $[R_\tau][R_{\tau\pi}]$ \ \\
\ $(133)(321)$ \ & & $\tilde{\nu}_{3L}$ & & & \ $0.02$ $[R_\tau][R_{D_s}(\tau\mu)]$ \\
\cline{3-3}
\ $(231)(212)$ \ & & $\tilde{\nu}_{2L}$ & & & \ $0.003$ $[R_\tau][R_\pi]$ \\
\ $(231)(221)$ \ & & $\tilde{\nu}_{2L}$ & & & \ $0.005$ $[R_\tau][R_{D^0}]$ \\
\hline
\end{tabular}
\caption{
The $2\sigma$ (95\% C.L.) bounds on R-parity violating couplings with the mediating sparticle masses
set to 100~GeV.  The indices on $\lambda$ have be reordered using the anti-symmetry in the
first two indices.   The rightmost column shows the product of the $2\sigma$ single-coupling bounds
listed in Table~\ref{SingleCouplingBounds}.
The observables that provide the individual constraints are shown in brackets.}
\label{BoundComparison}
\end{table}

\endgroup
\end{turnpage}

\newpage
\begin{table}[h]
\begin{tabular}{|c|l|l|}
\hline
\ Coupling\ \ &\ $2\sigma$ bound\ \ &\ Observable\ \ \\
\hline 
$\lambda_{12k}$ & \ $0.03\,\tilde{e}_{kR}$ & \ $V_{ud}$ \\
$\lambda_{13k}$ & \ $0.05\,\tilde{e}_{kR}$ \ ($0.03\,\tilde{e}_{kR}{}^*$)\ \ & \ $R_\tau=\Gamma(\tau\rightarrow e\nu_e\nu_\tau)/\Gamma(\tau\rightarrow \mu\nu_\mu\nu_\tau)$ \ \\
$\lambda_{23k}$ & \ $0.05\,\tilde{e}_{kR}$ & \ $R_\tau$ \ \\
\hline
$\lambda'_{111}$ & \ $7\times 10^{-4}\, \tilde{q}^2\tilde{g}^{1/2}$\ \ & \ $0\nu\beta\beta({}^{76}\mathrm{Ge}$)\\
$\lambda'_{11k}$ & \ $0.03\,\tilde{d}_{kR}$ & \ $V_{ud}$, $R_\pi=\Gamma(\pi\rightarrow e\nu_e)/\Gamma(\pi\rightarrow\mu\nu_\mu)$ \ \\
$\lambda'_{12k}$ & \ $0.2\,\tilde{d}_{kR}$ & \ $R_{D^+}=\Gamma(D^+\rightarrow \mu^+\nu_\mu\overline{K^0})/\Gamma(D^+\rightarrow e^+\nu_e\overline{K^0})$ \ \\
$\lambda'_{1j1}$ & \ $0.03\,\tilde{u}_{jL}$ & \ $Q_W({}^{133}\mathrm{Cs})$ \\
$\lambda'_{1j2}$ & \ $0.28\,\tilde{u}_{jL}$ & \ $A_\mathrm{FB}^s$ \\
$\lambda'_{21k}$ & \ $0.06\,\tilde{d}_{kR}$ \ ($0.04\,\tilde{d}_{kR}{}^*$) & \ $R_\pi$ \ ($R_{\tau\pi}=\Gamma(\tau\rightarrow\pi\nu_\tau)/\Gamma(\pi\rightarrow\mu\nu_\mu)$) \\
$\lambda'_{22k}$ & \ $0.1\,\tilde{d}_{kR}$ & \ $R_{D^0}=\Gamma(D^0\rightarrow \mu^+\nu_\mu K^-)/\Gamma(D^0\rightarrow e^+\nu_e K^-)$ \\
$\lambda'_{23k}$ & \ $0.45$ $(m_{\tilde{d}_{kR}}\!=100~\mathrm{GeV})$ \ & \ $R_\mu^Z = \Gamma(Z\rightarrow\mathrm{had})/\Gamma(Z\rightarrow \mu^+\mu^-)$ \ \\
$\lambda'_{31k}$ & \ $0.06\,\tilde{d}_{kR}$ \ ($0.08\,\tilde{d}_{kR}{}^*$)\ & \ $R_{\tau\pi}$ \\
$\lambda'_{32k}$ & \ $0.3\,\tilde{d}_{kR}$ & \ $R_{D_s}(\tau\mu)=\Gamma(D_s\rightarrow\tau\nu_\tau)/\Gamma(D_s\rightarrow\mu\nu_\mu)$ \\
$\lambda'_{33k}$ & \ $0.58$ $(m_{\tilde{d}_{kR}}\!=100~\mathrm{GeV})$ \ & \ $R_\tau^Z = \Gamma(Z\rightarrow\mathrm{had})/\Gamma(Z\rightarrow \tau^+\tau^-)$ \ \\
\hline
\end{tabular}
\caption{The $2\sigma$ bounds on single R-parity violating couplings from a variety of sources.
The notation follows that of Ref.~\cite{Barbier:2004ez} with the sparticle symbol
representing the sparticle mass divided by 100~GeV.
Only the current best bounds are shown.
The numbers have been updated from those given in Table 6.1 of Ref.~\cite{Barbier:2004ez} (page 110)
taking into account the most recent data available in the Review of Particle Properties \cite{Amsler:2008zzb}
and elsewhere.
In particular,
the bound on $\lambda'_{1j1}$ from the weak charge of Cesium-133 uses
the result of Ref.~\cite{Porsev:2009pr}.
The bound on $\lambda'_{111}$ from neutrinoless double beta decay uses the result of Ref.~\cite{KlapdorKleingrothaus:2006ff}, utilizing the nuclear matrix elements calculated in Ref.~\cite{Hirsch:1995ek}.
It does not account for the pion-exchange contribution discussed in Ref.~\cite{Faessler:1996ph}.
The bounds on $\lambda_{13k}$, $\lambda'_{21k}$, and $\lambda'_{31k}$ inside parentheses with asterisks are what they would be if the preliminary $\tau$-decay data from Babar \cite{Swagato} are taken into account. 
They are not used to calculate the numbers in the rightmost column of Table~\ref{BoundComparison}.
The bounds based on LEP data, namely those
on $\lambda'_{1j2}$, $\lambda'_{23k}$, and $\lambda'_{33k}$, have not been updated.
The bounds on $\lambda'_{23k}$ and $\lambda'_{33k}$ are from loop effects and do not scale linearly with the squark mass. To rescale to squark masses other than $m_{\tilde{k}_R}\!=100\,\mathrm{GeV}$,
see Ref.~\cite{Lebedev:1999vc}.
The detailed derivation of these bounds is presented in Ref.~\cite{KT2}, except for
the bound on $\lambda'_{111}$ from neutrinoless double beta decay which will be discussed separately in
Ref.~\cite{KT3}.}
\label{SingleCouplingBounds}
\end{table}


\end{document}